\documentclass[superscriptaddress,secnumarabic,
amssymb,amsmath,nobibnotes,aps,prd,showkeys,showpacs,nofootinbib]{revtex4}%
\usepackage{graphicx}
\usepackage{epsf}
\usepackage{bm}
\usepackage{amsmath}
\usepackage{amsfonts}
\usepackage{amssymb}
\usepackage{epstopdf}
\usepackage{natbib}%
\providecommand{\U}[1]{\protect\rule{.1in}{.1in}}
\newcommand{\be}{\begin{equation}}
\newcommand{\ee}{\end{equation}}

\newcommand{\mincir}{\raise
-3.truept\hbox{\rlap{\hbox{$\sim$}}\raise4.truept\hbox{$<$}\ }}
\newcommand{\magcir}{\raise
-3.truept\hbox{\rlap{\hbox{$\sim$}}\raise4.truept\hbox{$>$}\ }}

\usepackage{color}

\begin{document}

\title{Evolving Cosmic Scenario in Modified Chaplygin Gas with Adiabatic Matter Creation}

\author{Subhra Bhattacharya}
\email{subhra.maths@presiuniv.ac.in}
\affiliation{Department of Mathematics, Presidency University, Kolkata-700073, India}
\author{Shibaji Halder}
\email{shibajihalderrrrm@gmail.com}
\affiliation{Department of Mathematics, Vidyasagar College, Kolkata-700006, India}
\author{Subenoy Chakraborty}
\email{schakraborty@math.jdvu.ac.in}
\affiliation{Department of Mathematics, Jadavpur University, Kolkata-700032, India}
\begin{abstract}

Modified Chaplygin Gas has been successful in describing the cosmic history of the universe from radiation to $\Lambda$CDM in standard cosmology, while particle creation mechanism in nonlinear thermodynamics can be used to explain inflation as well as late time acceleration. The present work is an attempt to explore the possibilities of obtaining an alternative explanations to cosmic evolution when modified Chaplygin gas is used in the context of particle creation mechanism.

\end{abstract}

\maketitle
\section{Introduction}
\label{Intro}

The relativistic second order thermodynamic theories of Muller \cite{mul}, Israel and Stewart \cite{is1}-\cite{his} and Pavon {\it et. al.} \cite{pav1},\cite{pav2} play crucial role in describing the evolution of the Universe as a sequence of dissipative processes. The theory proposes that deviations from equilibrium described by bulk stress, heat flow and shear stress can be treated as independent dynamical variables bounded by average molecular speed thereby ensuring causality. In a homogeneous and isotropic FRW universe the bulk viscous pressure is the only possible mechanism for dissipative processes. The bulk viscous pressure can be attributed to particle number changing processes in an expanding universe \cite{zel}-\cite{zim2} or it might due to coupling of the different components of the cosmic fluids \cite{ud}-\cite{sch}. Particle creation mechanism driving bulk viscous pressure has been extensively used to describe the dynamics and evolution of the early universe including early inflation and current accelerated expansion \cite{sc1}. Particle creation has also been related to emergent universe \cite{sc2}.

In the present work we shall consider the cosmological implications of bulk viscous pressure due to particle creation mechanism in a universe with matter described by the Modified Chaplygin gas (MCG). Thermodynamically we shall concentrate on an isentropic universe, i.e. we shall envisage the universe as an open thermodynamic system, where entropy per particle under the mechanism of particle creation is constant \cite{pri,cal}, although there will be entropy creation due to particle creation driven phase space change. The viscous effects shall be described by the truncated Muller-Israel-Stewart (MIS) type theory.

The reason for considering Modified Chaplygin Gas (MCG) as the cosmic fluid is that it provides a unified Dark Matter-Dark energy manifestations in a single fluid. MCG is an exotic fluid with EoS 
\begin{equation}
p=A\rho-\frac{B}{\rho^{\alpha}},~0\leq\alpha\leq 1.\label{eos}
\end{equation}
This EoS show that MCG can accommodate a radiation dominated universe for $A=1/3$ at high density to one with negative pressure at low density for the current accelerating universe \cite{deb1,bena}. For $\alpha=-1,~B=1+A$ one can get the $\Lambda$CDM universe, while for $B=0,$ the EoS describes a perfect fluid, i.e. a quintessence model. Several works on MCG has established its consistency as a cosmic fluid \cite{wu,bed,costa,deb2,tha}.

Using a FRW model of the universe with viscous effects described by the MIS theory in an isentropic universe with particle creation, we shall find expressions for the Hubble parameter in terms of particle creation rate. Corresponding relevant cosmological parameters like the scale factor, deceleration parameter and energy density are evaluated along with basic thermodynamic variables like fluid temperature $T$ and particle number density $n$ in terms of the particle creation rate. Using a single phenomenological choice of the particle creation rate we shall then show a unified cosmic evolution starting from early accelerated expansion to a late time accelerated one. Further we could successfully connect the particle creation rate, in MCG to an increasing entropy in the de-Sitter phase. Finally we shall relate the particle creation mechanism to Hawking radiation \cite{haw}.

The paper is organised as follows: Section 2 deals with basic conditions for bulk viscous cosmology related to particle creation mechanism. Bulk viscous FRW universe with MCG has been presented in section  3. Section 4 shows a comparison of the present result for specific choices of the particle creation rate with recent observations. Section 5 shows the scalar field description corresponding to Model 3. The entropy production for the present context has been derived in section 6. Section 7 describes interrelation between particle creation process with Hawking radiation. Finally the paper ends with a brief discussion in section 8.

\section{Bulk viscous universe with particle creation: Non equilibrium M-I-S type thermodynamic theory }

The energy momentum tensor of a relativistic fluid with bulk viscosity as the only dissipative phenomenon is given by
\begin{equation}
T^{\mu\nu}=(\rho+p+\Pi) u^{\mu}u^{\nu}+(p+\Pi)g^{\mu\nu}\label{em}
\end{equation}
where $u^{\mu}$ is the 4 velocity, $\rho$ is the energy density, $p$ is the thermodynamic pressure and $\Pi$ is the bulk viscous pressure. Considering the second order non equilibrium thermodynamics, the entropy flow vector $S^{\mu}$ is characterised by the equation \cite{is2}
\begin{equation}
S^{\mu}=sN^{\mu}-\frac{\tau \Pi^{2}}{2\zeta T}u^{\mu}\label{entropy}
\end{equation}
where $N^{\mu}=nu^{\mu}$ is the particle flow vector with $n$ being the particle number density, $s,$ the entropy per particle, $\tau$ is the relaxation time, $T$ is the temperature of the fluid and $\zeta$ is the coefficient of bulk viscosity. 

Now we consider a scenario where the non vanishing bulk viscous pressure is due to a change in fluid number density, which is characterised by the particle production rate $\Gamma=\frac{\dot{N}}{N}, ~N=na^{3}$ being the number of particles in co-moving volume $a^{3}.$ For $\Gamma>0$ we get particle creation while $\Gamma<0$ usually means particle annihilation.  The varying particle number density will cause a change of phase creating a entropy production density which will be given by:
\begin{equation}
S^{\mu}_{;\mu}=-\frac{\Pi}{T}\left[3H+\frac{\tau}{s}\dot{\Pi}+\frac{1}{2}\Pi T\left(\frac{\tau}{\zeta T}u^{\mu}\right)_{;\mu}+\varepsilon\frac{n\Gamma}{\Pi}\right]\label{eprod}
\end{equation}
$\varepsilon$ being the chemical potential. For the validity of second law of thermodynamics we must have $S^{\mu}_{;\mu}=\frac{\Pi^{2}}{\zeta T}\geq 0.$ This gives the following non linear differential equation for bulk viscosity $\Pi$ \cite{zim2}
\begin{equation}
\Pi^{2}\left[1+\frac{1}{2}T\left(\frac{\tau}{\zeta T}u^{\mu}\right)_{;\mu}\right]+\tau\Pi\dot{\Pi}+3H\zeta\Pi=-\zeta\varepsilon n\Gamma.\label{pi}
\end{equation} 
Thus any deviation from equilibrium is characterized by the bulk viscous pressure $\Pi$ in the presence of particle creation $\Gamma,$ further the above equation asserts the existence of a single causal theory even with particle creation processes taken into account. 

Given the existence of particle creation $\Gamma,$ the conservation equations are modified as
\begin{equation}
N^{\mu}_{;\mu}=n\Gamma ;~~T^{\mu\nu}_{;\nu}=0 
\end{equation} which gives
\begin{align}
&\dot{n}+3Hn=n\Gamma \label{consv1}\\
\text{and}~&\dot{\rho}+3H(\rho+p+\Pi)=0\label{consv2}
\end{align} with $\dot{n}=n_{,\mu}u^{\mu}.$ Comparing equations (\ref{consv1}) and (\ref{consv2}) with the Gibb's relation 
\begin{equation}
Tds=d\left(\frac{\rho}{n}\right)+pd\left(\frac{1}{n}\right)\label{gibb}
\end{equation}
one can get 
\begin{equation}
nT\dot{s}=-3H\Pi-(p+\rho)\Gamma.\label{entropy1}
\end{equation}
Considering that the pressure $p$ and density $\rho$ are related to the thermodynamic variables $n$ and $T$ by the equations $p=p(n,T)$ and $\rho=\rho(n,T)$ and using the conservation equations (\ref{consv1}) and (\ref{consv2}) together with 
\begin{equation}
\frac{\partial\rho}{\partial n}=\frac{p+\rho}{n}-\frac{T}{n}\frac{dp}{dT}\label{rhovariat}
\end{equation} one can obtain the temperature evolution equation as
\begin{equation}
\frac{\dot{T}}{T}=-3H\left[\frac{\partial p/\partial T}{\partial \rho/\partial T}+\frac{\Pi}{T\partial \rho/\partial T}\right]+\Gamma\left[\frac{\partial p/\partial T}{\partial \rho/\partial T}-\frac{p+\rho}{T\partial \rho/\partial T}\right]\label{t}
\end{equation}
Alternatively using (\ref{entropy1}) the above relation can be written as
\begin{equation}
\frac{\dot{T}}{T}=-(3H-\Gamma)\frac{\partial p/\partial T}{\partial \rho/\partial T}+\frac{n\dot{s}}{\partial \rho/\partial T}.\label{tgamma}
\end{equation}
Thus it is easily observed that particle production affects the temperature with an effective viscous pressure $\Pi$ together with a direct coupling.

Considering isentropic particle production characterised by constant entropy $\dot{s}=0$ the viscous pressure can be obtained directly in terms of particle production rate as
\begin{equation}
\Pi=-\frac{\Gamma}{3H}(p+\rho).\label{pie}
\end{equation}
From the above we can get a cosmic fluid characterised by changing particle number density. Also the variation of the fluid temperature is now given by
\begin{equation}
\frac{\dot{T}}{T}=-(3H-\Gamma)\frac{\partial p}{\partial \rho}.\label{tisen}
\end{equation}
Further from (\ref{consv1}) for isentropic particle production the evolution of $n$ is given by
\begin{equation}
\frac{\dot{n}}{n}=-(3H-\Gamma)\label{n}
\end{equation}

\section{Bulk viscous FRW universe with MCG as cosmic fluid}
\label{model}

We consider a spatially flat FRW model of the homogeneous universe as an open thermodynamic system with metric  
\begin{equation}
ds^{2}=-dt^{2}+a^{2}(t)[dr^{2}+r^{2}(d\theta^{2}+\sin^{2}\theta d\phi^{2})]\label{metric}
\end{equation}
Since we shall consider the above metric in the context of non-equilibrium thermodynamics driven by particle creation mechanism, the corresponding cosmic fluid with dissipation $\Pi$ will  have the field equations given by 
\begin{equation}
3H^{2}=\kappa\rho;~~~~\dot{H}=-\frac{\kappa}{2}(\rho+p+\Pi)\label{fe}
\end{equation} with $H$ the Hubble parameter and $\kappa=8\pi G$ is the Einstein's gravitation constant. Considering MCG as the cosmic fluid with EoS given by (\ref{eos}) together with (\ref{consv2}) one can obtain the energy density of the fluid as
\begin{equation}
\rho^{\alpha+1}=\frac{B}{A+1}+\frac{C}{A+1}a^{-3\mu}e^{\mu\int\Gamma dt}\label{rho}
\end{equation}
with $\mu=(A+1)(\alpha+1)$ and $C$ the constant of integration. Then using (\ref{fe}) above the Hubble parameter is obtained as (choosing $\kappa=1$)
\begin{equation}
H=\frac{1}{\sqrt{3}}\left[\frac{B}{A+1}+\frac{C}{A+1}a^{-3\mu}e^{\mu\int\Gamma dt}\right]^{\frac{1}{2\alpha+2}}.\label{h}
\end{equation}
Using $\dot{H}=-\frac{1}{2}\left(1-\frac{\Gamma}{3H}\right)(p+\rho)$ the deceleration parameter $q=-\left(1+\frac{\dot{H}}{H^{2}}\right)$ is obtained as
\begin{equation}
q=-1+\frac{3}{2}\left(1-\frac{\Gamma}{3H}\right)\frac{C(A+1)}{Ba^{3\mu}e^{-\mu\int\Gamma dt}+C}.\label{dp}
\end{equation}
Also using (\ref{tisen}) and (\ref{n}) the thermodynamic variables $T$ and $n$ can be obtained as
\begin{equation}
T=T_{c}\rho^{-\alpha}\left(\frac{C}{A+1}\right)^{\frac{\mu-1}{\mu}}a^{-3(\mu-1)}e^{(\mu-1)\int\Gamma dt}\label{tm}
\end{equation}
and
\begin{equation}
n=n_{c}C^{\frac{1}{\mu}}a^{-3}e^{\int \Gamma dt}\label{nm}
\end{equation}
where $n_{c},~T_{c}$ are constants of integration.

 Equations (\ref{rho})-(\ref{nm}) clearly show that the relevant dynamical and thermodynamic parameters all vary as the particle creation rate $\Gamma.$ Thus we need specific choices of $\Gamma$ to trace the history of the cosmic evolution. In the absence of any known form of the particle creation parameter we make three phenomenological choices of the particle creation indicator $\Gamma.$ First two being functions of the scale factor $a,$ while the third being a function of the Hubble parameter $H.$
 \begin{align*}
\Gamma&=15\beta H[1-a\tanh(10-12a)]&\textbf{Model 1}\\
\Gamma& = 3\beta H(1+a^{m+1}),&\textbf{Model 2}\\
\Gamma&=3\beta H\frac{(A+1)H^{2\alpha+2}-B/3^{\alpha+1}}{(A+1)H_{0}^{2\alpha+2}-B/3^{\alpha+1}}&\textbf{Model 3}
\end{align*}
where $\beta$ is some positive constant and $m$ is any real number $(\neq -1).$ Model 1 and 2 are phenomenological choices based on the scale factor $a.$ (Recently a similar form as Model 1 was considered in \cite{nun} to describe Phantom behaviour through particle creation). In a universe dominated by bulk viscosity with perfect fluid, it is usually found that particle creation rate is proportional to the energy density \cite{gunzig}. Thus one can at least speculate that an universe dominated by bulk viscosity in the presence of an exotic fluid, like MCG, the particle creation rate will have some functional dependence on the energy density, hence we chose Model 3 as a function of the Hubble parameter $H.$

It is known that in a simplified model of the homogeneous and isotropic universe the dynamical variable is the scale factor $a(t),$ quantities that determine the time dependence of $a(t)$ must also determine the aspects of universe's evolution. Taylor series expansion of $a(t)$ about present time reveals two such important terms as the Hubble parameter $H$ and the deceleration parameter $q$. The deceleration parameter is the first non linear correction of the expansion and is indicative of cosmic expansion. Essentially expansion rate of the universe is given by the Hubble parameter, such that a $H>0$ indicates an expanding universe while $q$ will indicate the time dependence of $H,$ such that an accelerating universe will have $q<0$ and decelerating universe will have $q>0.$ Hence corresponding to the above three phenomenological anasatze we shall consider the dynamic and thermal evolution of the universe by tracing the evolution of the parameters like the Hubble parameter $H,$ deceleration parameter $q,$ the energy density $\rho,$ particle number density $n$ and fluid temperature $T.$

\vspace{-1.5em}
\subsection{Model 1:$\Gamma=15\beta H[1-a\tanh(10-12a)]$ }

Using the expression for $\Gamma$ as given above in the equations (\ref{rho})-(\ref{nm}) we obtain the following expressions for $\rho,~H,~q,~n~\text{and}~T$
\begin{align}
\rho&=\left[\frac{B+Ca^{15\mu\beta-3\mu}\cosh^{\frac{5\mu\beta}{4}}(10-12a)}{A+1}\right]^{\frac{1}{\alpha+1}}\label{rho1}\\
H&=\frac{1}{\sqrt{3}}\left[\frac{B+Ca^{15\mu\beta-3\mu}\cosh^{\frac{5\mu\beta}{4}}(10-12a)}{A+1}\right]^{\frac{1}{2\alpha+2}}\label{h1}\\
q&=-1+\frac{3C(A+1)\left[ 1-5\beta(1-a\tanh(10-12a))\right]}{2Ba^{3\mu(1-5\beta)}\cosh^{-\frac{5\mu\beta}{4}}(10-12a)+2C}\label{q1}\\
n&=n_{c}C^{\frac{1}{\mu}}a^{15\beta-3}\cosh^{\frac{5\beta}{4}}(10-12a)\label{n1}\\
T&=T_{c}\left(\frac{C}{A+1}\right)^{\frac{\mu-1}{\mu}}\rho^{-\alpha}a^{(3\mu-3)(5\beta-1)}\cosh^{\frac{5\beta(\mu-1)}{4}}(10-12a)\label{t1}
\end{align}

The above physical quantities are graphically evaluated using the following values of the parameters $A=\frac{1}{3},~B=3.5,~\alpha=0.5,~\beta=0.12~\text{and}~C=2$ \cite{sad, costa}.
\begin{figure}[htb]
\centering
  \begin{tabular}{@{}ccc@{}}
\includegraphics[width= 0.3\linewidth]{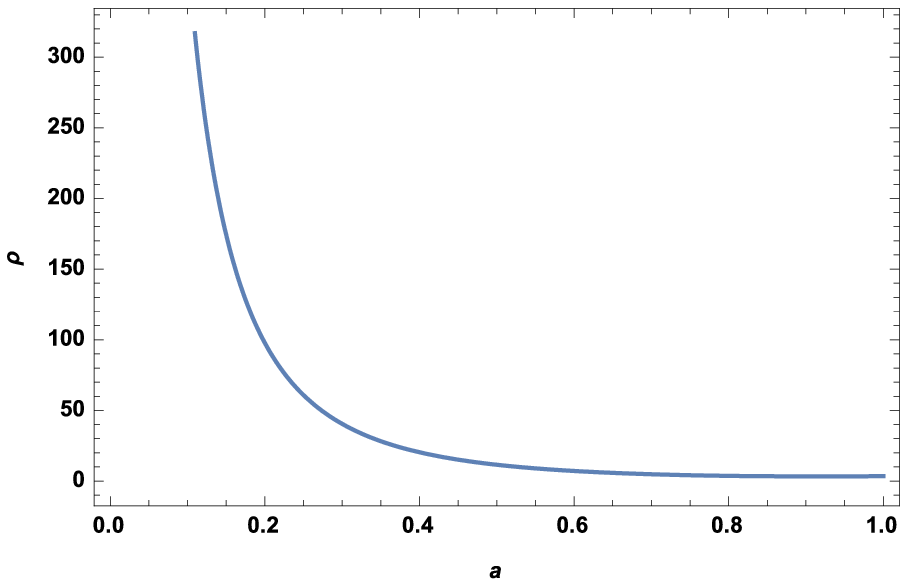}&
\includegraphics[width= 0.3\linewidth]{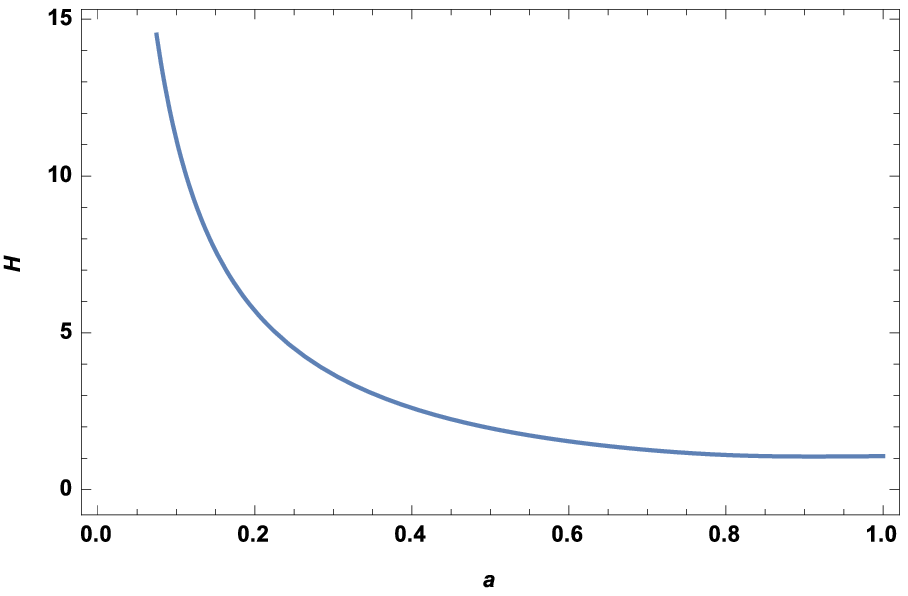}&
\includegraphics[width= 0.3\linewidth]{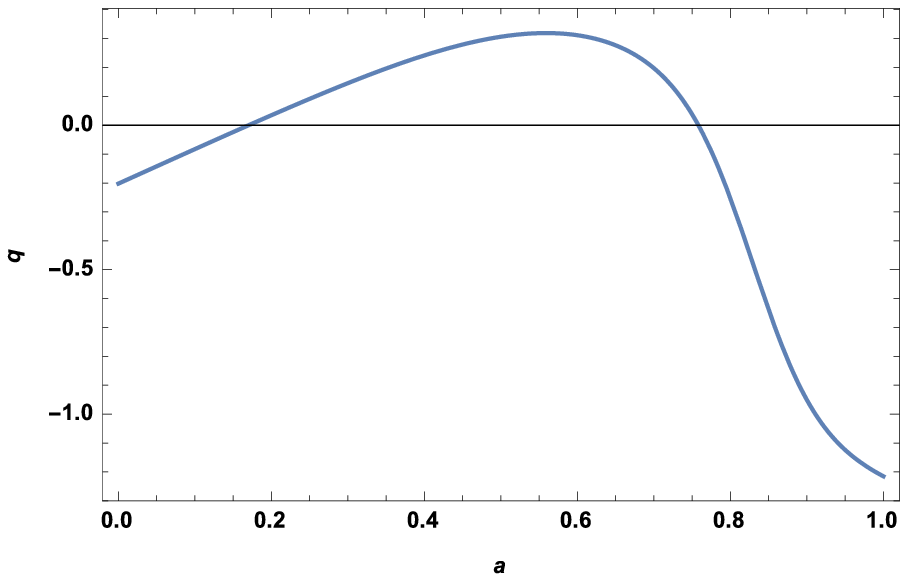}\\
\includegraphics[width= 0.3\linewidth]{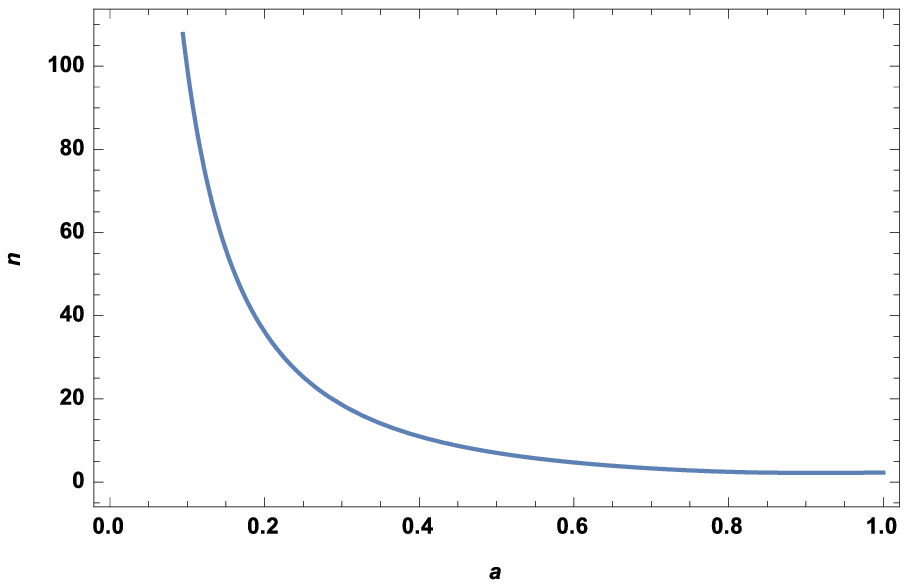}&
\includegraphics[width= 0.3\linewidth]{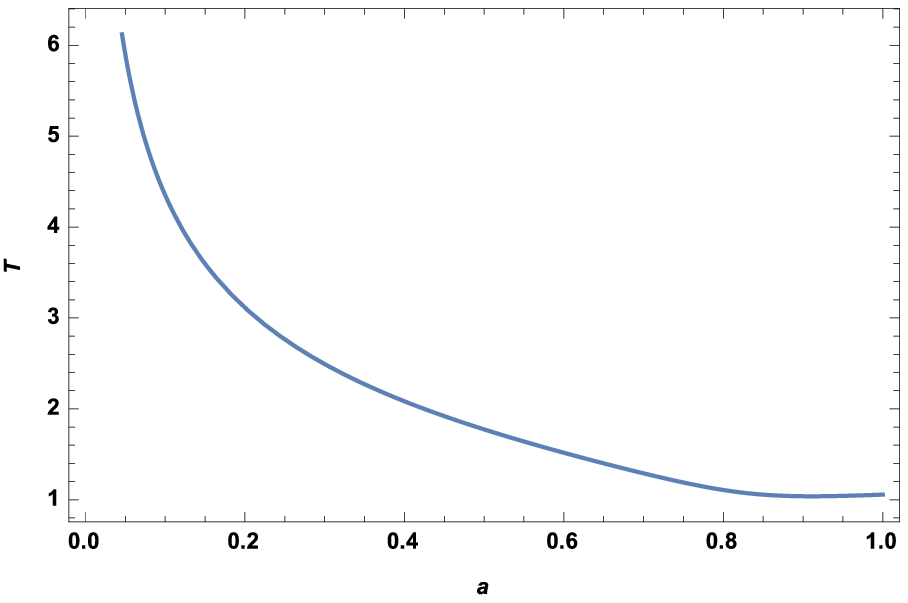}&
\end{tabular}
\caption{ The qualitative evolution of $\rho,~H,~q,~n,~T$ for $\Gamma=15\beta H[1-a\tanh(10-12a)]$ } 
\end{figure}

\subsection{Model 2: $\Gamma=3\beta H(1+a^{m+1})$}

As before we can obtain the following expressions for $\rho,~H,~q,~n~\text{and}~T$ corresponding to the above $\Gamma$ as
\begin{align}
\rho&=\left[\frac{B+Ca^{3\mu(\beta-1)}e^{\frac{3\mu\beta a^{m+1}}{m+1}}}{A+1}\right]^{\frac{1}{\alpha+1}}\label{rho2}\\
H&=\frac{1}{\sqrt{3}}\left[\frac{B+Ca^{3\mu(\beta-1)}e^{\frac{3\mu\beta a^{m+1}}{m+1}}}{A+1}\right]^{\frac{1}{2\alpha+2}}\label{h2}\\
q&=-1+\frac{3C(A+1)\left[1-\beta(1+a^{m+1})\right]}{2Ba^{3\mu(1-\beta)}e^{-\frac{3\mu\beta a^{m+1}}{m+1}}+2C}\label{q2}\\
n&=n_{c}C^{\frac{1}{\mu}}a^{3(\beta-1)}e^{\frac{3\beta a^{m+1}}{m+1}}\label{n2}\\
T&=T_{c}\left(\frac{C}{A+1}\right)^{\frac{\mu-1}{\mu}}\rho^{-\alpha}a^{3(\mu-1)(\beta-1)}e^{\frac{3\beta(\mu-1)a^{m+1}}{m+1}}\label{t2}
\end{align}

The above physical parameters are graphically presented in Fig 2 for $m=-2$ and using the previous values for the parameters $A,~B,~\text{and}~\alpha,$ while $\beta=0.08~\text{and}~C=1.$
\begin{figure}[htb]
\centering
  \begin{tabular}{@{}ccc@{}}
\includegraphics[width= 0.3\linewidth]{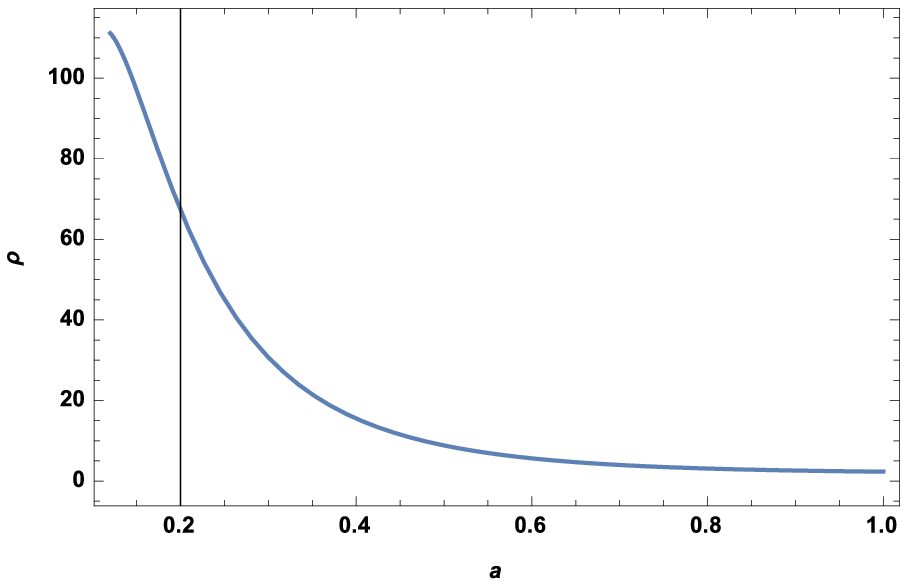}&
\includegraphics[width= 0.3\linewidth]{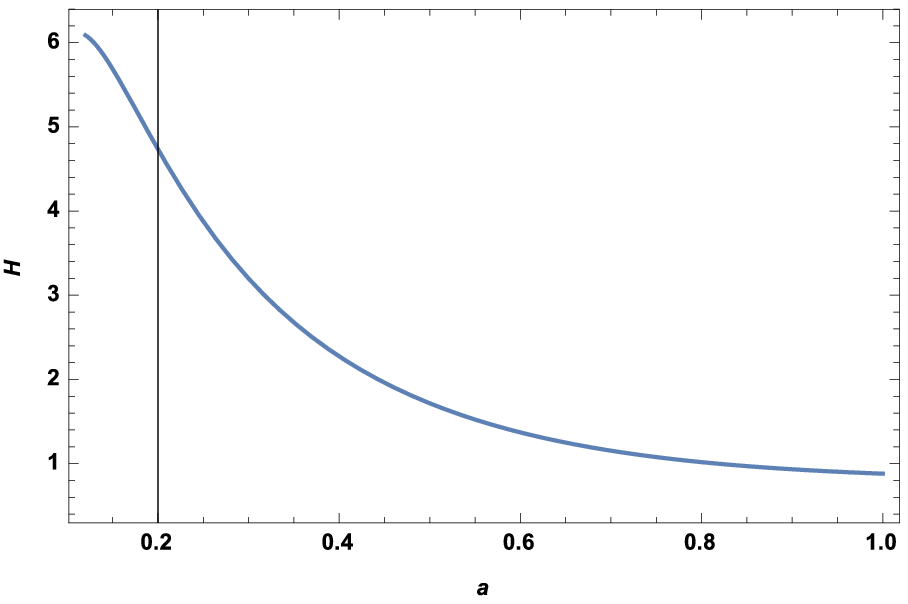}&
\includegraphics[width= 0.3\linewidth]{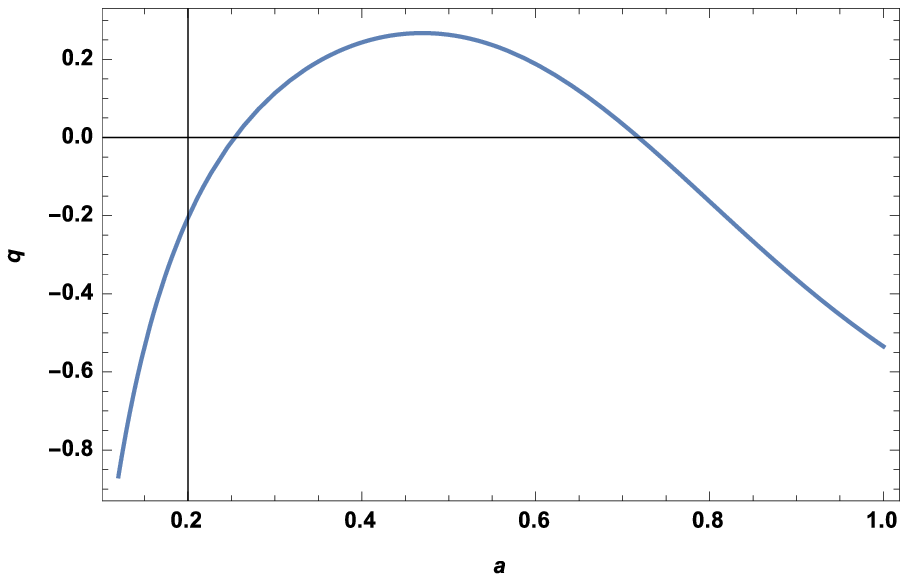}\\
\includegraphics[width= 0.3\linewidth]{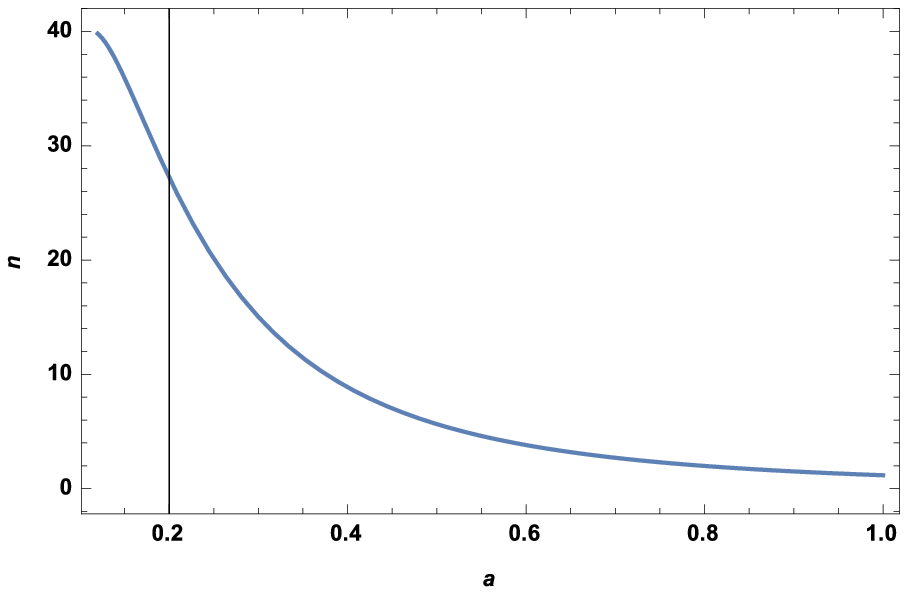}&
\includegraphics[width= 0.3\linewidth]{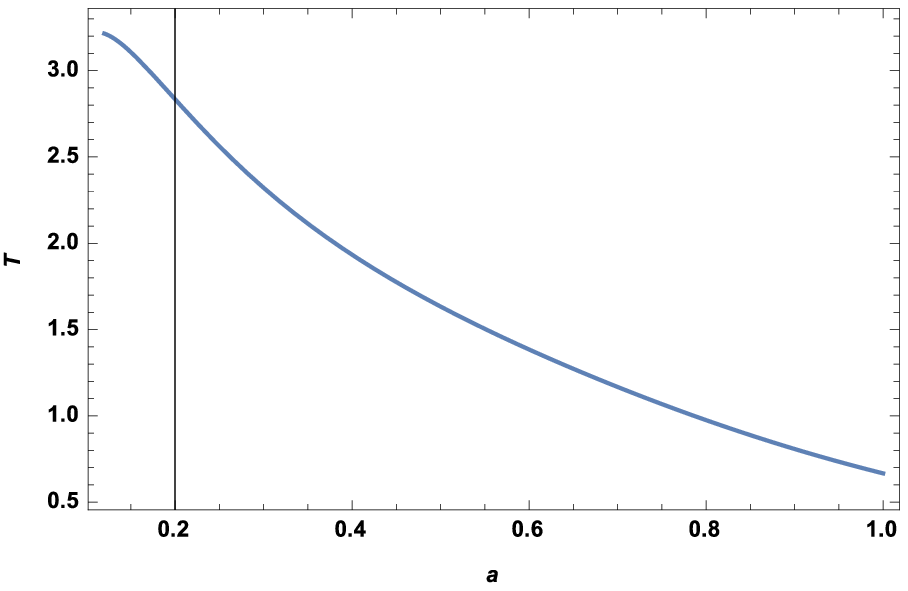}&
\end{tabular}
\caption{ The qualitative evolution of $\rho, ~H,~q,~n,~T$ for $\Gamma=3\beta H(1+a^{m+1})$} 
\end{figure}

\subsection{Model 3: $\Gamma=3\beta H\frac{(A+1)H^{2\alpha+2}-B/3^{\alpha+1}}{(A+1)H_{0}^{2\alpha+2}-B/3^{\alpha+1}}$}

As before we can obtain the following expressions for $\rho,~H,~q,~n~\text{and}~T$ corresponding to the above $\Gamma$
\begin{align}
\rho&=\left[\frac{\rho_{0}^{\alpha+1}}{r}+\frac{B}{A+1}\left(1-\frac{1}{r}\right)\right]^{1/(\alpha+1)}\label{rho3}\\
H&=\left(\frac{1}{A+1}\right)^{1/(2\alpha+2)}\left[\frac{y_{0}}{r}+\frac{B}{3^{\alpha+1}}\right]^{1/2\alpha+2}\label{h3}\\
q&=-1+\frac{3y_{0}(A+1)(r-\beta)}{2r(y_{0}+rB/3^{\alpha+1})}\label{q3}\\
n&=\frac{n_{0}}{r^{1/\mu}}\label{n3}\\
T&=T_{0}\left(\frac{\rho_{0}}{\rho}\right)^{\alpha}r^{-1+\frac{1}{\mu}}\label{t3}
\end{align}
where $r=\beta+(1-\beta)\left(\frac{a}{a_{0}}\right)^{3\mu},~y_{0}=(A+1)H_{0}^{2\alpha+2}-B/3^{\alpha+1}$ and $H=H_{0}$ at $a=a_{0}.$
 
The above physical parameters are graphically presented in Fig 3 using the previous values for the parameters $A,~B,~\text{and}~\alpha,~\beta$ while $a_{0}=0.3.$
\begin{figure}[htb]
\centering
  \begin{tabular}{@{}ccc@{}}
\includegraphics[width= 0.3\linewidth]{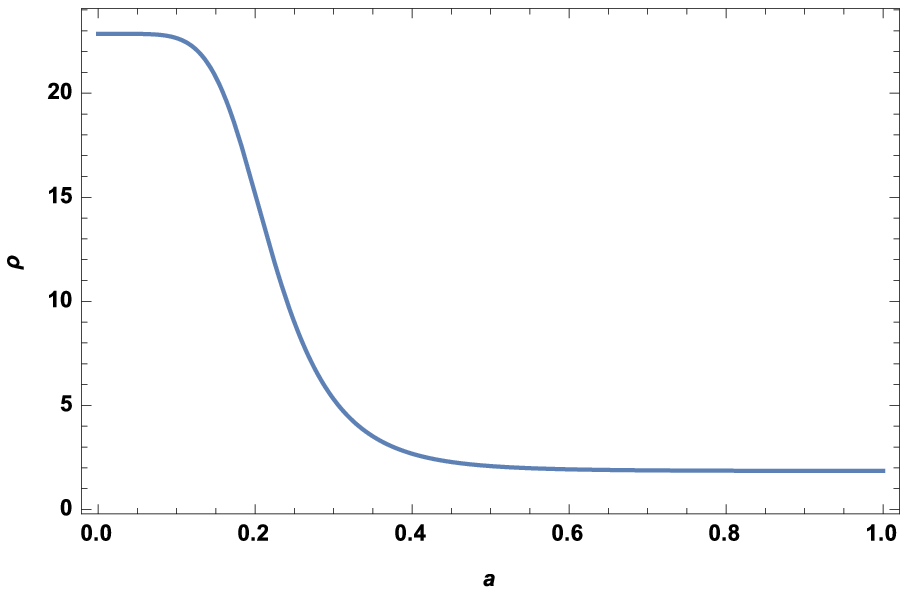}&
\includegraphics[width= 0.3\linewidth]{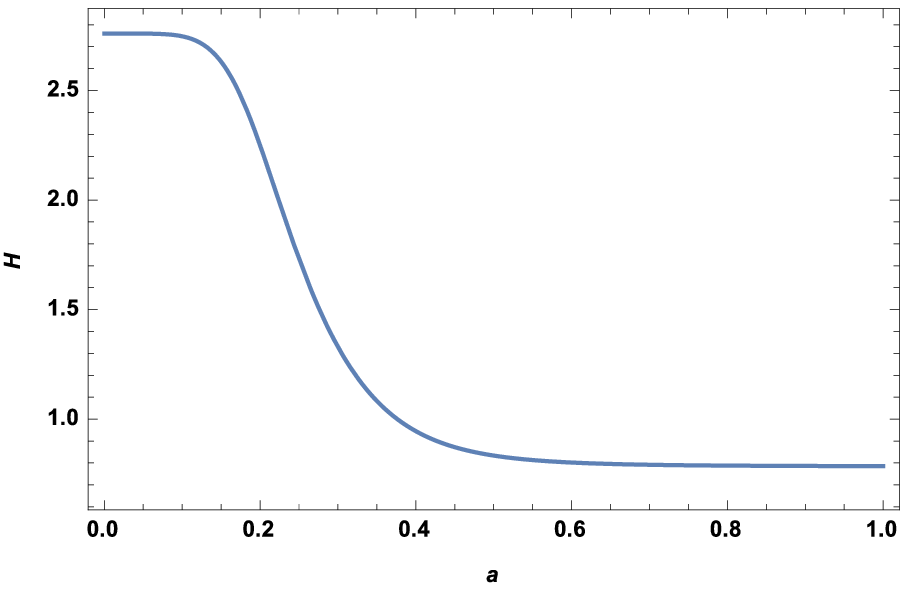}&
\includegraphics[width= 0.3\linewidth]{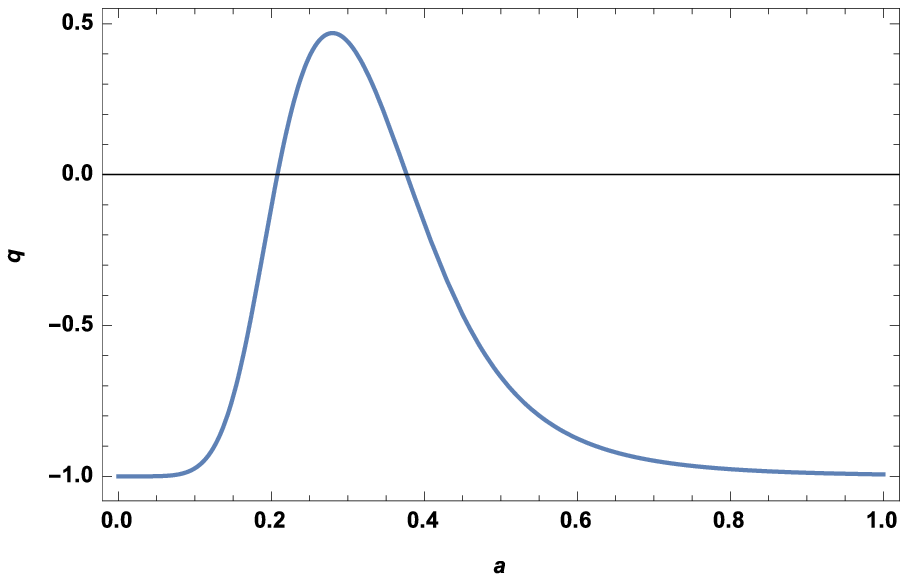}\\
\includegraphics[width= 0.3\linewidth]{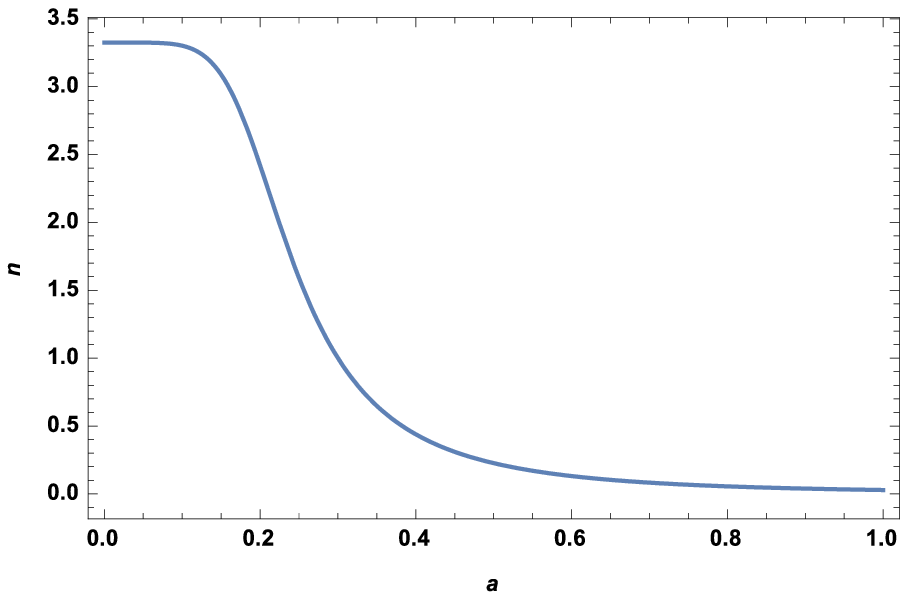}&
\includegraphics[width= 0.3\linewidth]{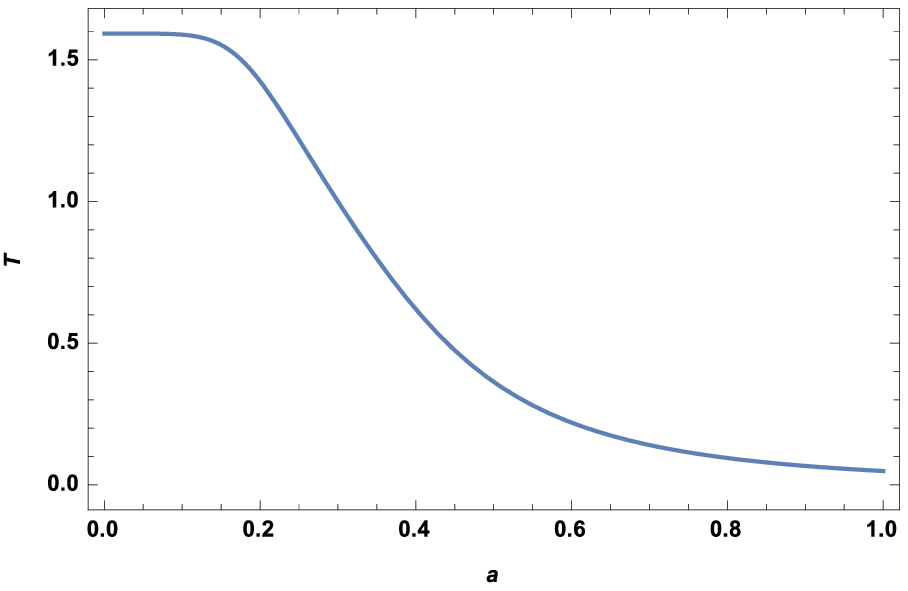}&
\end{tabular}
\caption{ The qualitative evolution of $\rho, ~H,~q,~n,~T$ for $\Gamma=3\beta H\frac{(A+1)H^{2\alpha+2}-B/3^{\alpha+1}}{(A+1)H_{0}^{2\alpha+2}-B/3^{\alpha+1}}$} 
\end{figure}

\section{Data comparison}
\label{data-analysis}

The Hubble parameter gives the expansion rate of the universe, and its time dependence can be measured using the deceleration parameter $q(t).$ In order to understand the kinematics of the cosmological evolution we consider a Taylor series expansion of the scale factor about the present time $t_{0}$ is given by 
\begin{equation}
a(t)=a(t_{0})\left[1+H_{0}(t-t_{0})-\frac{1}{2!}q_{0}H_{0}^{2}(t-t_{0})^{2}+\frac{1}{3!}j_{0}H_{0}^{3}(t-t_{0})^{3}+\frac{1}{4!s_{0}}H_{0}^{4}(t-t_{0})^{4}+\frac{1}{5!}l_{0}H_{0}^{5}(t-t_{0})^{5}+O((t-t_{0})^{6})\right]
\end{equation}
where $H_{0},~q_{0},~j_{0},~s_{0},~l_{0}$ are higher order derivatives of the scale factor considered at the present time and are more commonly known as the cosmographic Hubble, deceleration, jerk, snap and lerk functions respectively \cite{vis}. Considering that the scale factor $a(t)$ is related to the redshift $z$ by the relation $\frac{a(t_{0})}{a(t)}=1+z$ one can obtain the deceleration parameter $q(z)$ as a power series in $z$ as \cite{gui}
\begin{equation}
q(z)=q_{0}+(-q_{0}-2q_{0}^{2}+j_{0})z+\frac{1}{2}(2q_{0}+8q_{0}^{2}+8q_{0}^{3}-7q_{0}j_{0}-4j_{0}-s_{0})z^{2}+O(z^{3})
\end{equation} 
\begin{center}
\begin{table}
\caption{ First figure compares the late time evolution of the deceleration parameter for model 1 and 2 against data set 1, 2 and 3. The second figure compares the combined early and late time evolution of the deceleration parameter for model 1 and 2 against the latest data sets 4, 5 and 6. $q_{1},~q_{2}$ and $q_{3}$ is corresponding to models 1, 2 and 3.} 
\begin{tabular}{l*{3}{c}|l*{2}{c}}\hline
Data&$q_{0}$&$j_{0}$&$s_{0}$&Data&$q_{0}$&$j_{0}$\\
\hline
Data1(192 SN 1a+GRB - CPL)&-0.90&3.93&-25.52&Data4(BAO data)&-0.764&1.774\\
Data2(192 SN 1a+GRB - Linear)&-0.75&2.21&-12.25&Data5(Union2.1 SN1a+BAO+$H(z)$)&-0.48&0.68\\
Data3(Union2 SN 1a+GRB +BAO+OHD)&-0.39&-4.925&-26.40&Data6(Union2.1 SN1a+BAO+GRB)&-0.6&0.7\\
\hline
\end{tabular}
\caption{ Data used for generating graphics} 
\end{table}
\end{center}
\begin{center}
\begin{figure}[htb]
  \begin{tabular}{@{}ccc@{}}
\includegraphics[width= 0.5\linewidth]{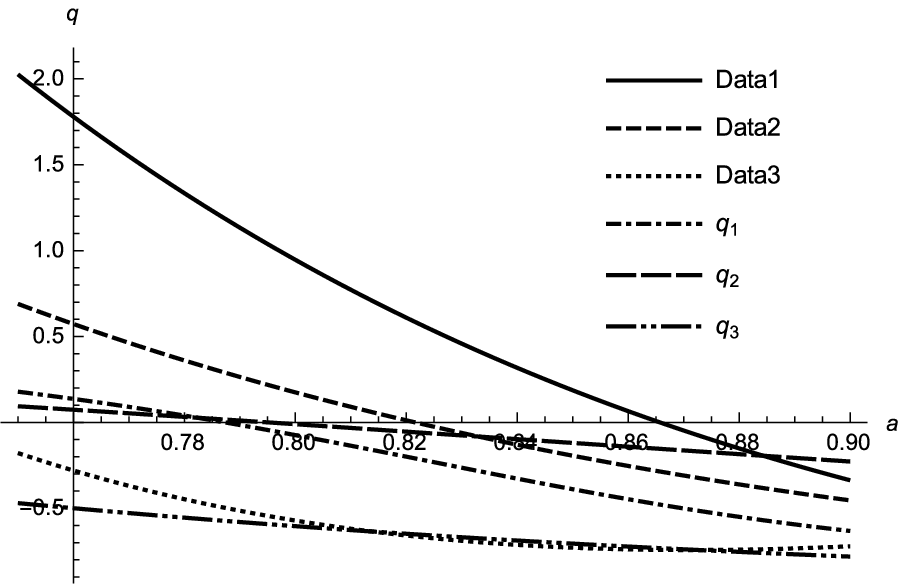}&
\includegraphics[width= 0.5\linewidth]{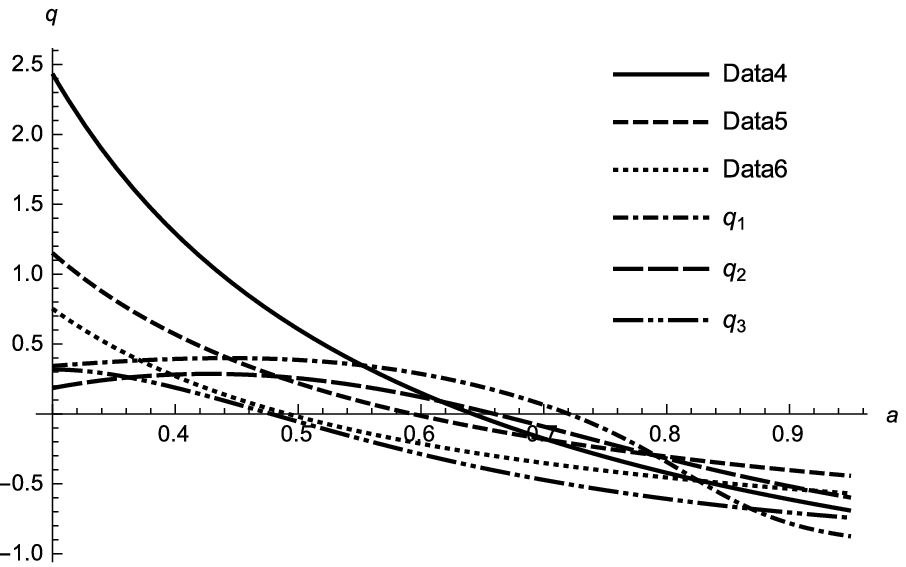}&
\end{tabular}
\end{figure}
\end{center}
Using the above expression for $q$ we have plotted the deceleration parameter corresponding to six data sets in Fig 4. Data 1 corresponds to 192 SN 1A and 69 GRB's with CPL parametrizations \cite{wang}, Data 2 corresponds to the same data set with linear parametrizations \cite{wang} and Data 3 corresponds to Union2+BAO+OHD+GRB's data \cite{xu} for low red shift range.  Data 4 corresponds to most recent high red-shift BAO data \cite{laz}, while Data 5 and Data 6 correspond to high red-shifts data from Union 2.1 compilation of SN 1a data with BAO and H(z) data, and SN 1a+GRB's+BAO data respectively \cite{demia}. (It may be noted that for data sets 4, 5 and 6 only first order terms were used because $s_{0}$ values were not constrained well in the literatures cited). Parameters used in the models are $A=1/3,~B=3.4,~\alpha=0.5,~C=2,~\beta=0.1~\text{and}~m=-2.$ 

From the left figure of Fig. 4 we can see that our model fits well with both Data 2 and Data 3, while from the right figure of Fig. 4 it is evident that our model accurately matched the data 4, 5 and 6 in the region $.5\leq a\leq .8$ with third model providing the best match.

\section{Field Theoretic Description  }
\label{sec-ftd}

In this section we shall address the process from the field theoretic view point. With the help of Model 3 we shall show the whole dynamical process as a evolution of scalar field $\phi$ having self interacting potential $V(\phi).$ We know that in terms of the scalar field, the energy density and thermodynamic pressure of the cosmic fluid is given by:
\begin{equation}
\rho=\frac{1}{2}\dot{\phi}^{2}+V(\phi)~~~~p_{tot}=p+\Pi=\frac{1}{2}\dot{\phi}^{2}-V(\phi).\label{phi}
\end{equation}
Now using the isentropic condition (\ref{pie}) together with the particle creation rate as used in Model 3 and the corresponding expression for Hubble parameter as given by (\ref{h3}) we can obtain
\begin{equation}
\phi=\phi_{s}+\frac{2}{\mu}\sqrt{\frac{\beta}{3y_{0}}}\frac{m}{\sqrt{1+m^{2}}}F\left[\sin^{-1}\sqrt{\frac{(A+1)H^{2\alpha+2}}{B/3^{\alpha+1}}}|\frac{m^{2}}{1+m^{2}}\right]
\end{equation}
where $F$ is the incomplete Elliptic integral of the first kind \cite{abram} and $m=\sqrt{\frac{B/3^{\alpha+1}}{y_{0}/\beta}}.$ Here scalar field $\phi$ always has a value greater than $\phi_{s}.$   Consequently
\begin{equation}
\left(\frac{a}{a_{0}}\right)^{3\mu}=\frac{y_{0}}{1-\beta}\left\lbrace\frac{B}{3^{\alpha+1}}\left[sn^{-1}\left(\frac{(\phi-\phi_{s})}{\frac{2}{\mu}\sqrt{\frac{\beta}{3y_{0}}}\frac{m}{\sqrt{1+m^{2}}}}|\frac{m^{2}}{1+m^{2}}\right)\right]^{2}-1\right\rbrace^{-1}-\frac{\beta}{1-\beta}
\end{equation}
 where $sn^{-1}$ is the equivalent inverse Jacobian Elliptic function. Now from (\ref{phi}) we know $2V(\phi)=\rho-p-\Pi,$ substituting the values of $\rho,~p,~\Pi$ in terms of the Hubble parameter $H$ as obtained in model 3, we get:
 \begin{equation}
 2V(\phi)=\nu H^{2\alpha}+\zeta H^{2}+\eta H^{2\alpha+4}.\label{vh}
\end{equation}
where 
\begin{align*}
\zeta&=\frac{B}{3^{\alpha}}\left(1-\frac{\beta}{y_{0}}\frac{B}{3^{\alpha+1}}\right)\\
\nu&=3\left(\frac{1-A}{1+A}-(2+A)\frac{\beta}{y_{0}}\frac{B}{3^{\alpha+1}}\right)\\
\eta&=\frac{3\beta}{y_{0}}(1+A)^{2}
\end{align*}   
Assuming $\psi=\frac{B}{3^{\alpha+1}(A+1)}\left[sn^{-1}\left(\frac{(\phi-\phi_{s})}{\frac{2}{\mu}\sqrt{\frac{\beta}{3y_{0}}}\frac{m}{\sqrt{1+m^{2}}}}|\frac{m^{2}}{1+m^{2}}\right)\right]^{2}$ as the modified scalar field, the corresponding modified field potential $2V(\phi)=\mathfrak{V}(\psi)$ is given by:
 \begin{equation}
 \mathfrak{V}(\psi)=\zeta\psi^{1-\delta}+\nu\psi^{\delta}+\eta\psi^{1+\delta}
 \end{equation}
where $\delta=\frac{1}{1+\alpha}.$ Thus, the potential is essentially a linear combination of different powers of the modified scalar field $\psi.$ Choosing $\alpha=0.5$ one can obtain 
 \begin{equation}
 \mathfrak{V}(\psi)=\zeta\psi^{1/3}+\nu\psi^{2/3}+\eta\psi^{5/3}.
 \end{equation}
 From the above relations the particle creation rate $\Gamma$ takes the form
 \begin{equation}
 \Gamma=\frac{3\beta}{y_{0}}\left[(A+1)\psi^{1+\frac{1}{2\alpha+2}}-\frac{B}{3^{\alpha+1}}\psi^{\frac{1}{2\alpha+2}}\right].
 \end{equation} Hence we obtain a continuous description of the scalar field evolution and corresponding potential for different cosmological epochs, establishing the correlation between scalar field and fluid evolution.
 
 For models 1 and 2 however, it is difficult to obtain explicit analytic expressions for $\phi$ and $V(\phi).$ They are thus solved by numerical methods and the corresponding scalar field evolution has been represented graphically in Figure 5. The leftmost figure corresponds to model 1, the middle to model 2 and the right most to model 3.
 
 \begin{figure}[htb]
\centering
  \begin{tabular}{@{}ccc@{}}
\includegraphics[width= 0.3\linewidth]{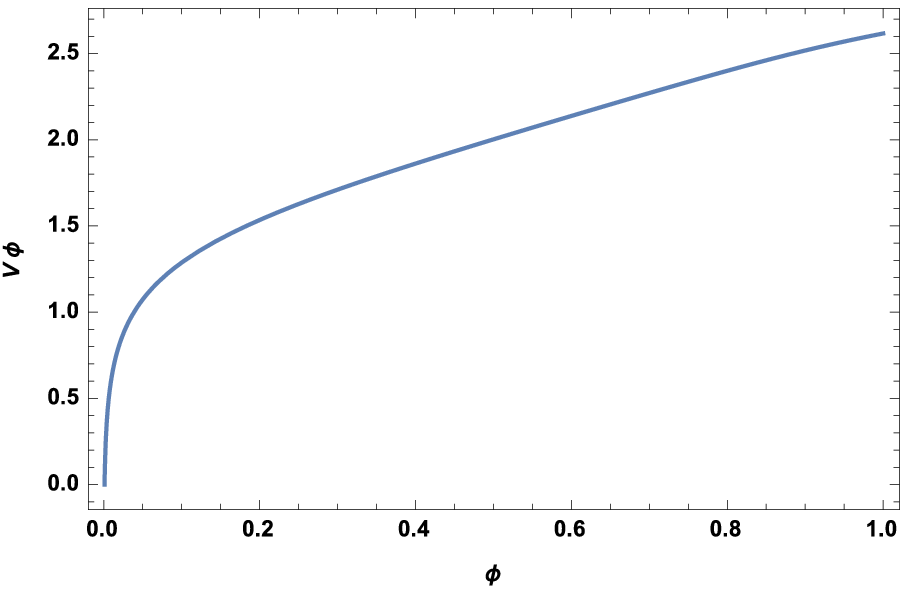}&
\includegraphics[width= 0.3\linewidth]{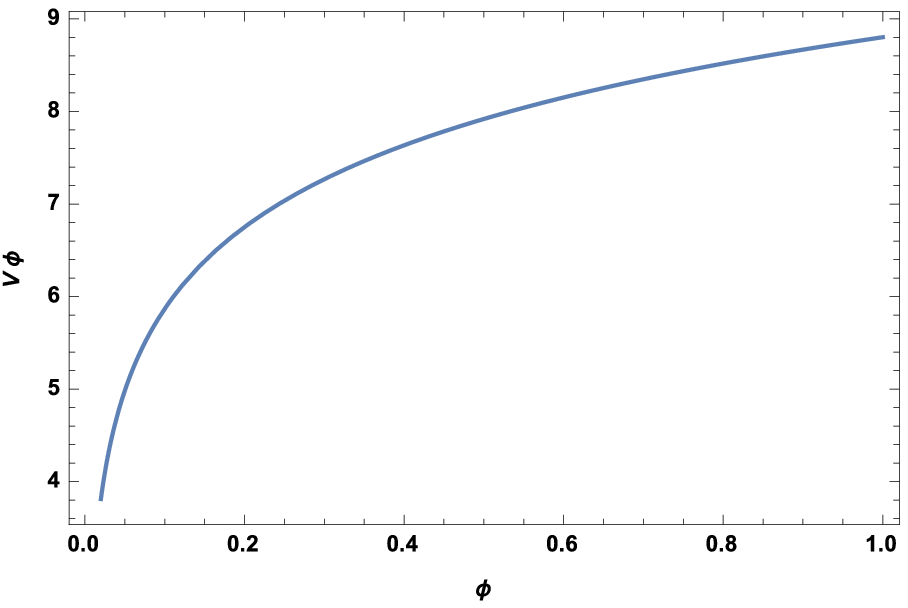}&
\includegraphics[width= 0.3\linewidth]{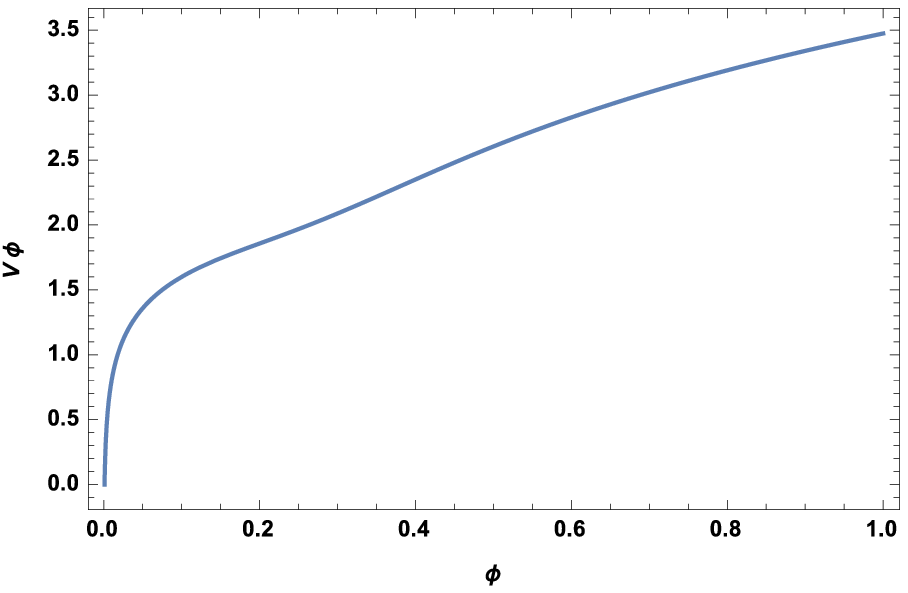}
\end{tabular}
\caption{ The graphical representation of $V(\phi)$ versus $\phi$ for the three models considered with parameter values same as figures 1, 2 and 3.} 
\end{figure}
 
 \section{Entropy production  }
\label{sec-ent}

Eliminating $\Pi$ from (\ref{entropy1}) and using (\ref{consv1}) and (\ref{pie}) we obtain the rate of change of entropy per particle as
\begin{equation}
nT\dot{s}=-\frac{\dot{n}}{n}(p+\rho)+\dot{\rho},
\end{equation}
after some algebraic manipulations which gives
\begin{equation}
TX^{\frac{\alpha}{\alpha+1}}\dot{s}=\frac{A+1}{\alpha+1}\left[\frac{dX}{dt}+n^{-1-\alpha}\frac{dY}{dt}\right]\label{sdot1}
\end{equation}
where 
\begin{equation}
X=\left(\frac{\rho}{n}\right)^{\alpha+1}~\text{and}~Y=\frac{B(\alpha+1)}{A+1}\log n-\frac{A}{A+1}\rho^{\alpha+1}.\label{substs}
\end{equation}
Also from (\ref{t}) using (\ref{consv1}) together with $\Pi=-\rho\left(\gamma+\frac{2}{3}\frac{\dot{H}}{H^{2}}\right)$ one can obtain
\begin{equation}
T=T_{i}e^{[\int_{n_{i}}^{n}\frac{v-u\gamma}{n}dn+\int_{\rho_{i}}^{\rho}\frac{u}{\rho}d\rho]}\label{tval}
\end{equation}
where $u$ and $v$ are the ratios, given by $u=\frac{\rho}{T\partial\rho/\partial T},~v=\frac{\partial p/\partial T}{\partial\rho/\partial T}$ and $\gamma=\frac{p+\rho}{\rho}.$ Subscript $i$ here refers to some initial time. Using this value of $T$ in (\ref{sdot1}) one obtains
\begin{equation}
\dot{s}=\left(\frac{n}{\rho}\right)^{\alpha}\frac{1}{T_{i}}e^{-[\int_{n_{i}}^{n}\frac{v-u\gamma}{n}dn+\int_{\rho_{i}}^{\rho}\frac{u}{\rho}d\rho]}\frac{A+1}{\alpha+1}\left[\frac{dX}{dt}+n^{-1-\alpha}\frac{dY}{dt}\right].\label{sdot2}
\end{equation}
For the inflationary phase with $\rho=\rho_{i}$ and $H=H_{i}$ the above expression for $\dot{s}$ simplifies to
\begin{equation}
\dot{s}=\frac{\gamma_{i}\rho_{i}}{nT_{i}}\left(\frac{n}{n_{i}}\right)^{-v_{i}+u_{i}\gamma_{i}}(3H_{i}-\Gamma)\label{sdot3}
\end{equation}
where $\gamma_{i}=A+1-\frac{B}{\rho_{i}^{\alpha+1}},~u_{i}=u|_{\rho_{i}},~v_{i}=v|_{\rho_{i}}.$ Corresponding to the radiation era with high density, we choose $\gamma_{i}=\frac{4}{3}$ and $u_{i}=1,~v_{i}=\frac{1}{3}.$ Using these in (\ref{sdot3}) one obtains
\begin{equation}
\dot{s}=\frac{4\rho_{i}}{3n_{i}T_{i}}(3H_{i}-\Gamma).\label{sdot4}
\end{equation}
Integrating the above we get
\begin{equation}
s=\frac{4\rho_{i}}{n_{i}T_{i}}\left[(t-t_{i})H_{i}-\frac{1}{3}\int_{t_{i}}^{t}\Gamma dt\right]+s(t_{i}).\label{s}
\end{equation}
From the above we find that $s$ has an additional dependence on the particle creation rate $\Gamma.$ Further one can find that the change of entropy in comoving volume given by $\mathcal{E}=sna^{3}$ using MCG as matter is similar to that obtained in \cite{zim2} with normal matter with bulk viscosity driven by particle creation. Using above result we can write an explicit expression for $\mathcal{E}$ as 
\begin{equation}
\mathcal{E}=\left\lbrace \frac{4\rho_{i}}{n_{i}T_{i}}\left[(t-t_{i})H_{i}-\frac{1}{3}\int_{t_{i}}^{t}\Gamma dt\right]+s(t_{i})\right\rbrace N_{i}e^{\int_{t_{i}}^{t}\Gamma dt}\label{v}
\end{equation}
which clearly shows that for $\Gamma\neq 0$ there is an exponential increase of comoving entropy in MCG in the de Sitter phase. Thus we can safely commit that with MCG as cosmic fluid, the results above are in complete agreement to the results obtained in \cite{zim2} corresponding to the normal fluid, and further we assert that the viscous pressure $\Pi$ is connected to an increasing particle number rather than with changing entropy per particle.

\section{Interrelation between particle creation with MCG and Hawking Radiation}
\label{section-hwr}
In the previous sections it is found that the dissipative phenomenon in the cosmic substratum leads to non-equilibrium thermodynamics with particle creation mechanism. Further the dissipative effect is only in the form of effective bulk viscous pressure due to homogeneity and isotropic nature of space-time. The present section is an attempt to show some inner relationship between particle creation rate and Hawking temperature.

In the context of universal thermodynamics, the process of Hawking radiation is just the opposite to black hole (BH) evaporation. The particles created just outside the event horizon in the BH evaporation escape outside towards asymptotic infinity, but in universal thermodynamics the created particles near the trapping horizon move inside the horizon. Also this flow of particles will be uniform in all directions due to the isotropic nature of space-time. Further at the beginning, the BH evaporation process is very slow, subsequently with the decrease of the BH size, the process becomes faster and faster, so that the temperature becomes larger and larger until quantum gravity effects become important due to the plank size of the BH. In universal thermodynamics, on the other hand, the universe at the beginning is of the Planck size and quantum gravity effects are important, then gradually with the expansion of the universe Hawking radiation comes into effect and the temperature gradually decreases.

There is another basic difference between the two processes namely, due to Hawking radiation in the context of BH there is a loss of energy and hence it is termed as BH evaporation, while due to Hawking radiation in universal thermodynamics, the universe gains energy and hence it should be referred as Hawking condensation. It is reasonable to assume that the Hawking radiation follows the Stephen-Boltzman radiation law \cite{haw,mod}.
\begin{equation}
P=\frac{dQ}{dt}=\sigma A_{\tau}T^{4}\label{st}
\end{equation}
where $\sigma=\frac{\pi^{2}\kappa_{B}^{2}}{60\hbar^{3}c^{2}}$ is Stephen-Boltzman constant, $A_{\tau}$ is the area of the bounding trapping horizon and $T=\frac{\hbar H}{2\pi\kappa_{B}}$ is the Hawking temperature. Using this heat variation in the first law of thermodynamics 
\begin{equation}
\dfrac{dQ}{dT}=\dfrac{d}{dt}(\rho V_{H})+p\dfrac{dV_{H}}{dt}=\sigma A_{H}T^{4} \label{flt}
\end{equation}
one gets
\begin{equation}
\dot{\rho}+3H(\rho+p)=3\sigma T^{4}\label{ec}
\end{equation}
where $A_{H}$ and $V_{H}$ are respectively the area and volume bounded by the horizon. The non-zero r.h.s. of the above equation gives some dissipation due to Hawking radiation. Thus comparing with energy-momentum conservation relation (\ref{consv2}), the effective bulk viscous pressure is given by 
\begin{equation}
\Pi=-\frac{\sigma T^{4}}{H}.
\end{equation}
Further using the relation(\ref{pie}) for isentropic process the particle creation rate is given by
\begin{equation}
\Gamma=\frac{\sigma T^{4}}{H(1+A-B(3H^{2})^{-(\alpha+1)})}
\end{equation}
Now using the value of $T,$ the Hawking temperature, as given above, we have
\begin{equation}
\Gamma=\frac{\sigma\hbar^{4} H^{2\alpha+5}}{(2\pi\kappa_{B})^{4}\lbrace(1+A)H^{2\alpha+2}-\frac{B}{3^{\alpha+1}}\rbrace}.
\end{equation}
Hence for large $H,~\Gamma\propto H^{3}.$ Since in the early phase of the evolution of the universe the MCG behaves as perfect fluid (with constant equation of state $p=A\rho$) we have
\begin{equation}
H^{-2}=H_{0}^{-2}+\left(\frac{a}{a_{0}}\right)^{3(A+1)}
\end{equation}
i.e. $H\sim H_{0}$ for $a\ll a_{0}$ while $H\sim a^{-\frac{3(1+A)}{2}}$ for $a\gg a_{0}.$ Thus one obtains the usual exponential expansion at the early phase and subsequently, the evolution follows the standard cosmology. On the other hand when $H$ is small (i.e. at the late phase of evolution) then instead of particle creation there will be particle annihilation and $\Gamma\propto H^{2\alpha+5}.$ This is not a physically realistic situation and hence Hawking type radiation is not possible at later phase of evolution.

\section{Discussions}
\label{discu}

In the above sections MCG model is considered as a candidate for describing cosmic evolution with dissipation in the form of bulk viscosity due to the mechanism of particle creation. For three different choices for the particle creation rate (as a function of Hubble parameter and scale factor) it is possible to show a complete cosmic scenario from inflation to present late time acceleration. Also the thermodynamic parameters namely density, temperature and particle number density are presented both analytically and graphically. The deceleration for the three models are compared with different observed results and it is found that the present models match with observed results for different ranges of the red shift parameter (or the scale factor). One of the three models (Model 3) has been shown equivalent to a a scalar field, with self interacting potential description. Analytic expression for the scalar field and the potential function has also been evaluated. Considering the particle creation mechanism in the context of non equilibrium thermodynamic prescription, the entropy production in the cosmic volume has been evaluated for the MCG model. We could show that in the radiation era, corresponding to high density the comoving entropy will have an exponential increase that is proportional to the particle creation rate $\Gamma.$ Finally an attempt has been made to find an analogy of the present model in Hawking radiation. Thus we can conclude that with MCG as cosmic fluid it is possible to obtain a continuous cosmic scenario in the presence of bulk viscosity due to particle creation. Unlike the cosmic scenario in the presence of a perfect fluid, that requires different particle creation parameters for different cosmic era \cite{sc1}, a cosmic scenario in the presence of MGC could be described by a single continuous choice of the particle create rate $\Gamma.$ It is only natural if one considers that MCG can as such accommodate an universe starting with inflationary scenario to the current accelerating one, while particle creation effects can explain both inflationary and late time acceleration of the universe. We also note that using equations (\ref{eos}) and (\ref{rho}) it is possible to evaluate the effective equation of state parameter $\omega_{eff}=\frac{\rho+p+\Pi}{\rho}$ for the models 1, 2 and 3 in terms of the scale factor $a.$ Defining constraints between the free parameters it is possible to obtain a qualitative behaviour of of $\omega_{eff}$ w.r.t. the scale factor. Figure 6 is the corresponding graphical representation. The figures show that at a very early time, our models can have an effective equation of state parameter $\omega_{eff}=-1$ thus describing an accelerated expansion, while at the present time one can obtain an effective equation of state having phantom attributes. 

In conclusion we state that the present work successfully describes the cosmic evolution using a single continuous particle creation model in Modified Chaplygin gas, incorporating the early and late time accelerated expansion of the universe. It remains to be seen whether such continuous models can describe in details the inflationary cosmology successfully, and can be considered as a future work.

\begin{figure}[htb]
\centering
  \begin{tabular}{@{}ccc@{}}
\includegraphics[width= 0.3\linewidth]{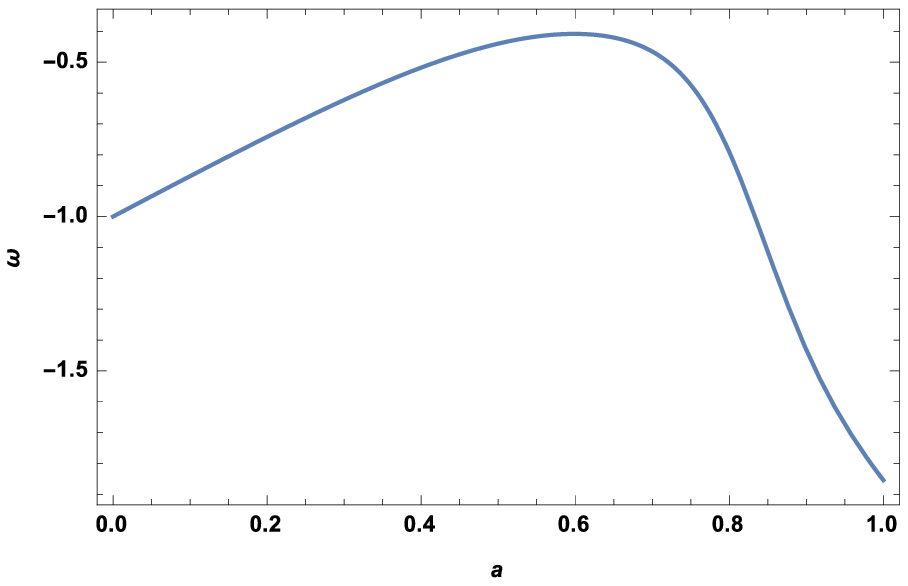}&
\includegraphics[width= 0.3\linewidth]{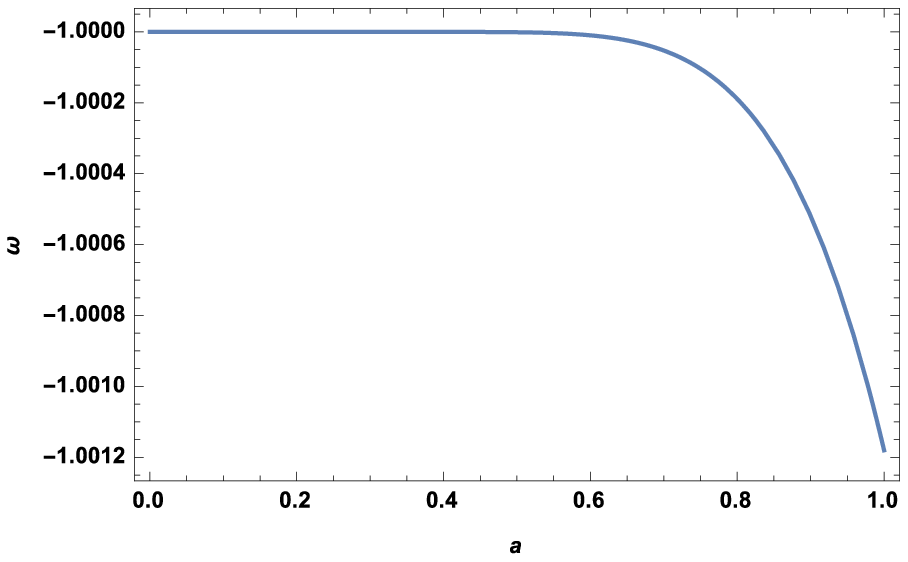}&
\includegraphics[width= 0.3\linewidth]{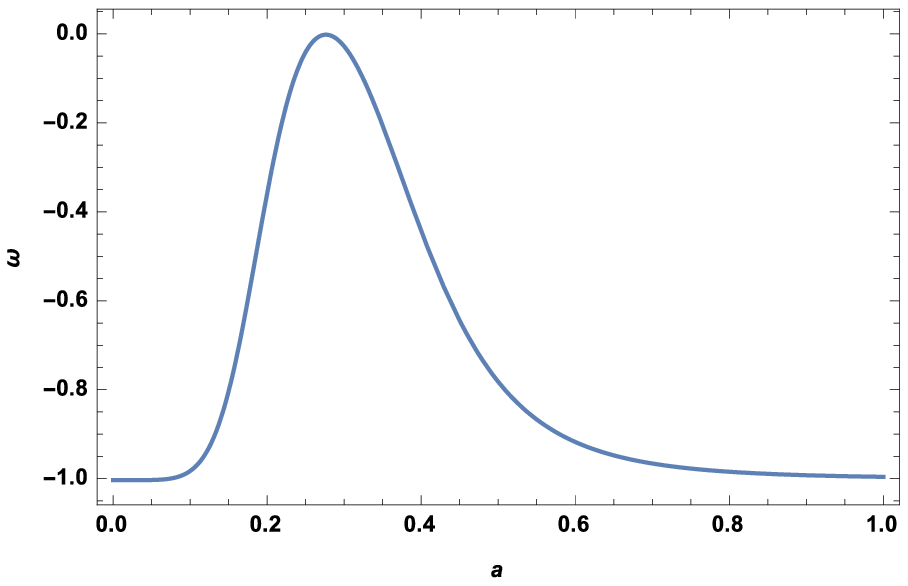}
\end{tabular}
\caption{ The graphical representation of the effective equation of state corresponding to the models 1, 2, 3 respectively. $A=\frac{1}{3},~B=3.34,~\alpha=0.5,~\beta= .2,~1.2,~.08$ respectively for Models 1, 2, 3.} 
\end{figure}

\section{Acknowledgments}
 SB acknowledges UGC's Faculty Recharge Programme. SC thanks IUCAA, Pune, India, for their warm hospitality while working on
this project.


\end{document}